\documentclass[aps,pra,showpacs,onecolumn]{revtex4}
\usepackage{graphicx}
\usepackage[inline]{showlabels}

\usepackage{amssymb}
\usepackage{amsmath}
\usepackage{braket}
\usepackage{xcolor}
\usepackage{hyperref}
\usepackage{color}
\usepackage{pbox}
\usepackage{float}
\usepackage{enumerate}

\newcommand{\redmark}[1] {\color{red}\textbf{#1}\color{black}\normalsize}

\usepackage[noend]{algorithm}
\usepackage[noend]{algcompatible}

\floatname{algorithm}{Algorithm}

\newcounter{nota}
\newcommand{\nota}[1]{ \stepcounter{nota}
                 \noindent{\bf Nota \Roman{nota}: }  }

\begin{document}

\title{Quantum state transfer performance of Heisenberg spin chains with site-dependent interactions designed using a generic genetic algorithm}

\author{Sofía Perón Santana$^{(1)}$}
\email{sofia.peron@mi.unc.edu.ar}
\author{Martín Domínguez$^{(2)}$}
\email{mdoming@famaf.unc.edu.ar}
\author{Omar Osenda$^{(1)}$ }
\email{osenda@famaf.unc.edu.ar}

\affiliation{
(1) Instituto de F\'isica Enrique Gaviola (CONICET-UNC) and Facultad de 
Matem\'atica, Astronom\'ia, F\'isica y Computaci\'on, Universidad Nacional de 
C\'ordoba,
Av. Medina Allende s/n, Ciudad Universitaria, CP:X5000HUA C\'ordoba, Argentina\\
(2) Facultad de 
Matem\'atica, Astronom\'ia, F\'isica y Computaci\'on, Universidad Nacional de 
C\'ordoba,
Av. Medina Allende s/n, Ciudad Universitaria, CP:X5000HUA C\'ordoba, Argentina .
}

\date{\today}

\begin{abstract}
Designing a good transfer channel for arbitrary quantum states in spin chains implies optimizing a cost function, usually the averaged fidelity of transmission. The fidelity of transmission measures how much the transferred state resembles the state prepared at the beginning of the transfer protocol. When averaged over all the possible initial states, the figure of merit quantifies the quality of the protocol. There are proposals for optimizing a given Hamiltonian to accomplish a particular task. The transfer of quantum states is one of them. In particular, we consider the design of Heisenberg spin chains using a genetic algorithm. This very efficient algorithm allows us to study different properties of Hamiltonians with good to excellent transfer ability. One apparent drawback of using a random search method is that it results in exchange coefficient strengths that change abruptly from site to site. Modifying the cost function, we obtain Hamiltonians with exchange coefficients varying smoothly along the chain length. Our results show that the smoothed Hamiltonians have the same, or less, transfer ability than the rough ones, and both kinds show similar robustness against static disorder. By studying the statistical properties of the eigenvalues of Hamiltonians with varying transfer abilities, we determine the ensemble of random matrices to which the spectra belong.
\end{abstract}

\maketitle

\section{introduction}\label{sec:introduction}

The lack of exact results, coupled with the increasing difficulty in implementing the literal versions of quantum algorithms or protocols, has led to a rising reliance on optimization methods to determine feasible options for different quantum information processing tasks \cite{Meister2023,Gustiani2023,Perrier2020}. These options do not provide the desirable results but satisfy other requirements. They are fast, efficient, or simply doable. 

Among the quantum information tasks, quantum state transfer (QST) \cite{Chang2023, Keele2022, Xie2023, Maleki2021, Serra2022, Mograby2021} is one of the more malleable to be studied from different perspectives. The transfer protocol simply requires an initial state prepared in a copy of a given physical system to travel to another copy. Both copies should be connected by more copies of the system or by a completely different one, depending on whether the whole system is homogeneous or hybrid. 

As mentioned above, studies on QST analyze it as resulting from the time evolution owed to an autonomous or controlled Hamiltonian \cite{Wang2016,Burgarth2010n,Yang2010,Heule2010,Stefanatos2019,Watanabe2010,Kostak2007,quantum-dot-chain,Li2018,nuclear-spin-chain,Loft2011,Banchi2011prl,Chapman2016,Kandel2019,Coden2020,Zhang2016,Gong2007,Murphy2010,Faroq2015}, as a SWAP operation between the sites where the quantum state must travel \cite{Maleki2021}, a succession of SWAPS between neighboring sites \cite{Kandel2021}, and an optimization problem. All these studies have lead to extensive terminology describing the quality of the transfer, such as perfect transfer (PT) 
\cite{Christandl2004,Christandl2005,Burgarth2005a,Burgarth2005b,Bayat2014}, pretty good transfer (PGT) \cite{Serra2022,Serra2024,Burgarth2006, Godsil2012a, Godsil2012b, Vinet2012,Kay2019, Kay2010, Banchi2017, vanBommel2010}, almost perfect transfer, or the workings of the protocol, such as conclusive and arbitrarily perfect QST \cite{Burgarth2005c}, etc. All the mentioned studies, along with those not cited,  highlight a thriving field. Yet, there are numerous problems without an adequate answer. 

One drawback machine learning methods, or machine learning-like methods \cite{Carleo,Zhang2018}, have is the difficulty often encountered in comparing or assessing the advantages of one solution over another, beyond the improvement on the value of the cost function \cite{Carleo}. At times, the difficulty arises because the solution is too expensive in computational time, and there are no practical ways to obtain many different solutions. In other cases, a physics-motivated procedure to improve or to understand the solutions is lacking.

The enhancement of QST through the modulation of some exchange coupling coefficients (ECC) for nearest neighbor interactions XX-like Hamiltonians was the subject of many studies some years ago \cite{Banchi2010,Zwick2015}. For the XX Hamiltonian, optimizing the transmission fidelity in terms of a reduced number of ECC is a (relatively) simple task because the eigenvalues and other quantities that enter into the expression of the transmission fidelity have simple analytical expressions. The physical traits associated with the enhancement are simple. The spectrum near the center of the band presents a reduced number of equally spaced eigenvalues, which results in the propagation of a "wave packet" with constant group velocity and without dispersion. 

Interestingly, optimizing the strength of the interactions of XXZ Hamiltonians or the distance between sites of anisotropic Hamiltonians with long-range interactions to enhance their QST ability also results in a portion of their spectra becoming equally spaced \cite{SerraPLA,FerronPS}. The optimization in these cases is numerical since the spectra and other quantities do not have analytically manageable expressions. Also worth mentioning is the condition that states that the spectrum of a given Hamiltonian must have its successive eigenvalues approximately differing by odd multiples of a fixed quantity, ensuring the appearance of pretty good QST. While the presence of PGT is appealing, more and more evidence shows that this scenario results in excessively long transmission times, strongly reducing its practical interest. The PGT scenario guarantees that the probability that an excitation transfer between two points of the communication channel, $P$, approaches the unity as closely as desired, simply by allowing sufficient time. Serra and collaborators show in References \cite{Serra2022} and \cite{Serra2024} that for an ample class of systems, if $\varepsilon=1-P$, then the transmission time $T_{\varepsilon}\sim 1/(\varepsilon)^f$, with $f$ a number such that $f\leq N/2$, where $N$ is the length of the transmission channel. Nevertheless, even for the PGT scenario, chains with site-dependent interactions outperform the homogeneous chains. 

Given the advances in experimental setups, where chains of qubits with modulable site-dependent interactions are accessible with present technology \cite{Kandel2021, Martins2017, Kostak2007, quantum-dot-chain, Li2018, Loft2011, Banchi2011prl, Chapman2016, Kandel2019, Baum2021}and that time-dependent control of even moderate chains remains elusive, the design of chains with time-independent site-dependent Hamiltonians continues to offer an attractive option. 

Since the probability of transmission, or the fidelity of transmission, is a complicated function of the eigenvalues and the eigenvectors of the chain Hamiltonian, and whose analytical expressions are unknown or unmanageable, it seems reasonable to use global optimization methods that do not rely on derivatives or other analytical techniques. In this paper, we employ a version of the well-known genetic algorithm to design the ECC of spin chains. 

Genetic Algorithms (GAs) simulate biological evolution to find optimal solutions. Starting with an initial solution pool, GAs iteratively create new generations by combining and modifying the best candidates. Termination occurs based on predefined criteria.

In GAs, solutions are represented as vectors of parameters subject to modification. Key stages include initialization, fast fitness function-based evaluation, diverse candidate selection, and crucial reproduction phases involving mutation and crossover.
Mutation involves randomly changing one or more genes on a chromosome to explore new solutions in the search space. Crossover combines parts of two parent chromosomes to create one or more daughter chromosomes, allowing the transfer of genetic information between individuals.

Termination criteria vary, such as a fixed time/iteration limit or finding an acceptable solution. GAs offer an adaptable approach for optimization, evolving solutions over generations.

 Genetic algorithms are biased random search algorithms \cite{genetico} where a set of rules chosen to evolve a population of subjects provide the bias. The evolution improves the {\em fitness} of the population, and the fitness is the cost function to be optimized, in our case, the fidelity of transmission of a spin chain. 

As we will show, the genetic algorithm offers several advantages over other methods for global optimization, namely a set of simple rules that produce chains with excellent transfer abilities (high values for the fidelity of transmission together with fast transmission), the running time to find the optimal solution scales as $N^2$, where $N$ is the length of the chain, for chains up to one-hundred sites. That the scaling law exponent is only two is fundamental because one of the problems in understanding the physical traits of the optimal chains is that as they are the result of a random search, it is necessary to count with many different solutions to ascertain which properties are general and which are particular to one solution. Another advantage of the genetic algorithm (GA) results from the possibility of changing the fitness function, which allows studying qualitatively different solutions. 

We organized the paper as follows. Section \ref{sec:methods} is devoted to the system Hamiltonian, the transmission protocol, and some details about the GA. Section \ref{sec:probability-fitness} presents some results about the transmission ability of chains designed using the GA with different fitness functions. In Section \ref{sec:static-disorder}, we analyze the robustness of the designed chains against static disorder. In Section \ref{sec:random-matrices}, we present an analysis of the statistical properties of the spectra of the Hamiltonians associated with the designed chains. Finally, we discuss our results and present our conclusions in Section 6.

\section{Spin chain Hamiltonian, transmission protocol and genetic algorithm}\label{sec:methods}

We consider spin chains with a Hamiltonian given by

\begin{equation}
\label{eq:general-hamiltonian}
H= -\sum_{i=1}^{N-1} \, J_i (\mathbf{s_i^x s_{i+1}^x} + \mathbf{s_i^y s_{i+1}^y} + \mathbf{s_i^z s_{i+1}^z} ),
\end{equation}

\noindent where the $ \mathbf{s_i^{\alpha}} = \sigma^{\alpha}$,  $\sigma^{\alpha}$ are the Pauli matrices, $N$ is the chain legtn or, equivalently, the number of spins, and $J_i$ are the exchange coupling coefficients.  The Hamiltonian in Eq. \eqref{eq:general-hamiltonian} commutes with the total magnetization in the z-direction, so, as it is customary, we will focus on the propagation of excitations on the one-excitation subspace. 

Using the computational basis, the quantum state that describes a single excitation  localized in the site $i$ of the chain/graph  is denoted by
\begin{equation}
|\mathbf{i}\rangle = |0\rangle_1 \otimes |0\rangle_2 \otimes \ldots |1\rangle_i \otimes |0\rangle_{i+1} \ldots|0\rangle_N,
\end{equation}
where $|0\rangle_j$ and $|1\rangle_j$ are the basis vectors of the Hilbert space associated with the site $j$, $1\leq j \leq N$, and $N$ is the number of sites in the system. 

In the simplest transmission protocol, the initial state of the system is  the quantum state of a single excitation localized at one site
\begin{equation}
|\psi(t=0)\rangle = |\mathbf{i}\rangle.
\end{equation}
If the Hamiltonian of the system is time-independent, then the time-dependent quantum state is given by
\begin{equation}
|\psi(t)\rangle = e^{-itH} |\psi(t=0)\rangle,
\end{equation}
where $H$ is the Hamiltonian and $\hbar=1$. The probability of transmission from site $i$ to site $j$ at time $t$ is given by
\begin{equation}
P_{i,j}^{(N)} = \left| \langle \mathbf{i} | e^{-itH} | \mathbf{j} \rangle  \right|^2. 
\end{equation}

For systems where the total magnetization in the $z$ direction commutes with the Hamiltonian, the probability of transmission reduces to
\begin{equation}\label{eq:transmission-probability}
P_{i,j}^{(N)} = \left| \langle \mathbf{i} | e^{-it h_N} | \mathbf{j} \rangle  \right|^2, 
\end{equation}
where $h_N$ is the one excitation block Hamiltonian, given by a $N\times N$ Hermitian matrix \cite{Serra2022}. 

Most commonly, the transmission occurs between the first and last spins of the chain, so the quantity of interest is the probability $P_{1, N}$. The average fidelity of transmission is given by
\begin{equation}
f_{av}= \frac12 + \frac{\sqrt{P_{1, N} } }{3} + \frac16 P_{1, N} .
\end{equation}

It is clear that both functions, $P_{1, N}$ and $f_{av}$ are complicated functions of the eigenvalues and eigenvectors of the Hamiltonian. We aim to maximize the probability $P_{1, N}$ at a chosen arrival time as a function of the ECC values. Because it depends on the product of the time and the Hamiltonian, selecting a particular value of the arrival time changes the scale of the ECC values proportionately. For convenience, we choose values for the arrival time as natural multiples of the chain length. 

To maximize $P_{1, N}$, we consider some monotonous functions that depend on it as the fitness characterizing a given chain with its corresponding ECC. The purpose of genetic algorithms consists of maximizing the fitness of a population of chains.

\begin{figure}[bt]
\includegraphics[width=0.7\linewidth]{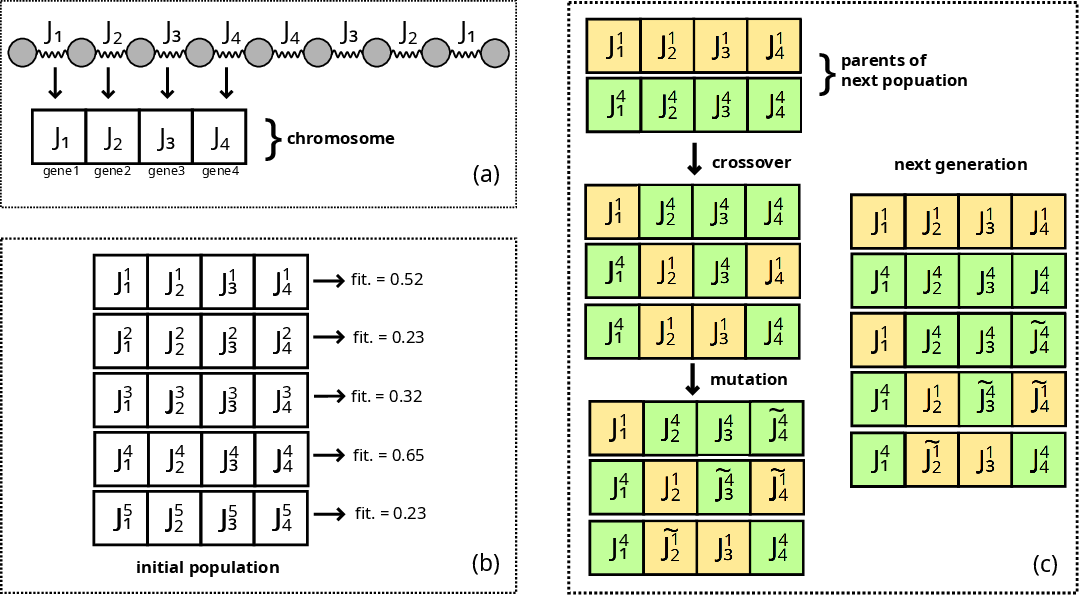}
\caption{The cartoon depicts the main elements of how a genetic algorithm improves, from generation through generation, the fitness of a population of spin chains. a) top: A centrosymmetric spin chain with ECCs $J_1.J_2, J_3.$ and $J_4$, {\em i.e.} an individual with four genes. Bottom: each $J_i$ belongs to a chromosome with four genes. b) The initial population of chains, with five individuals. Sorting randomly between allowed limiting values, each individual has genes $J_i^k$, where $i=1,2,3, 4 $  is the index numbering the genes, and $k=1,2,3,4$ and 5 is the supra index numbering the individuals. For each individual, the algorithm calculates its fitness. b) From top to bottom. Top:  The algorithm picks the fittest parents according to the value of the corresponding hyperparameter. Then, the algorithm eliminates those individuals who do not belong to the parent group. Middle: The algorithm generates the offspring through the crossover. That is, their chromosome combines the parent's genes. Note the color code that shows what gene corresponds to which parent. After that, the crossover has generated enough offspring to substitute those individuals eliminated. Bottom: some genes of the new individuals can mutate, which leads its value $J_l^m$  to another one, $\tilde{J}_l^m$. After the mutations, the algorithm evaluates the fitness of the individuals of the next generation and iterates until it meets the stopping conditions.}\label{fig:cartoon}
\end{figure}

The cartoon in Figure 1 depicts in its panels all the elements intervening in the genetic algorithm employed to obtain the majority of the results in the paper. Panel a) of the cartoon shows a centrosymmetric chain with two different ECCs, each coefficient a gene of the chain's chromosome. A population of chains is a set of chains, each chain with its own chromosome and corresponding fitness, as shown in b). Note that each individual of the population has its own chromosome. The algorithm selects several parents between the fittest individuals of the population and proceeds to try to increase the fitness of the next generation. The offspring resulting from the mating of the parents will replace the less fit individuals in the new generation. The offspring results from the crossover of the genes of the parents. Panel c) shows the three individuals resulting from the crossover, each with a combination of the parent's genes. After the crossover, there is a chance that the genes of the new individuals suffer a mutation, which changes their genes randomly. Finally, the fittest parents and their offspring conform to the new generation of the population. The algorithm iterates until achieving a prefixed tolerance or until the number of generations reaches a maximum value conveniently chosen.

To work, the packages that implement different versions of the genetic algorithms need several parameters called hyperparameters since they do not enter into the set of parameters of the problem under study.

\section{The chains obtained using different fitness functions} \label{sec:probability-fitness}

As said in the previous Section, the genetic algorithm needs several parameters that tune its behavior. For each Figure showing results, we will list the parameters employed. Appendix A has a detailed account of the effect of each parameter on the algorithm. A complete implementation of the algorithm employed to obtain the results shown in this Section can be found in \cite{SofiPS}.

In this Section, we include results obtained using two different fitness functions. The first one is the obvious choice of the probability of transmission as the fitness,

\begin{equation}\label{eq:fit1}
fit_1 = P_{1, N}(T),
\end{equation}
while the second one is given by
\begin{equation}\label{eq:fit2}
fit_2 = P_{1, N}(T) \times \left[ \gamma + \beta \exp \left(-\sum_{i=2}^{N-1}(J_i-J_{i-1})^2 )/N^2\right) \right],
\end{equation}

\noindent where $\gamma +\beta=1$.
There is no simple reason to justify our choice for $fit_2$. In any case, the justification comes from the results found using it.  

\begin{figure}[bt]
\includegraphics[width=0.4\linewidth]{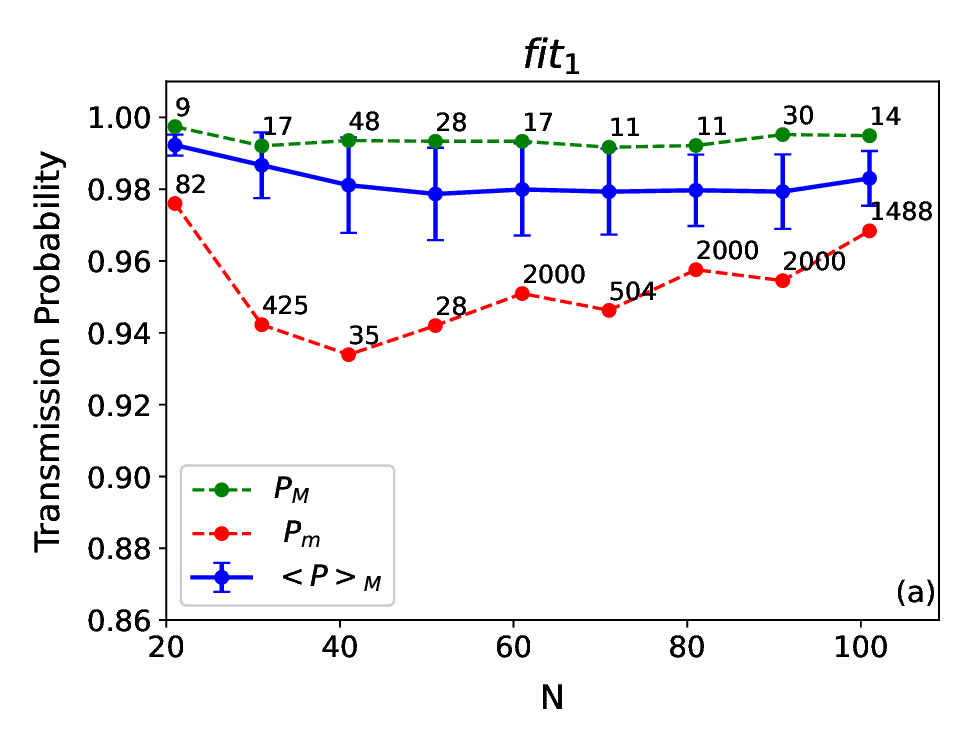}
\includegraphics[width=0.4\linewidth]{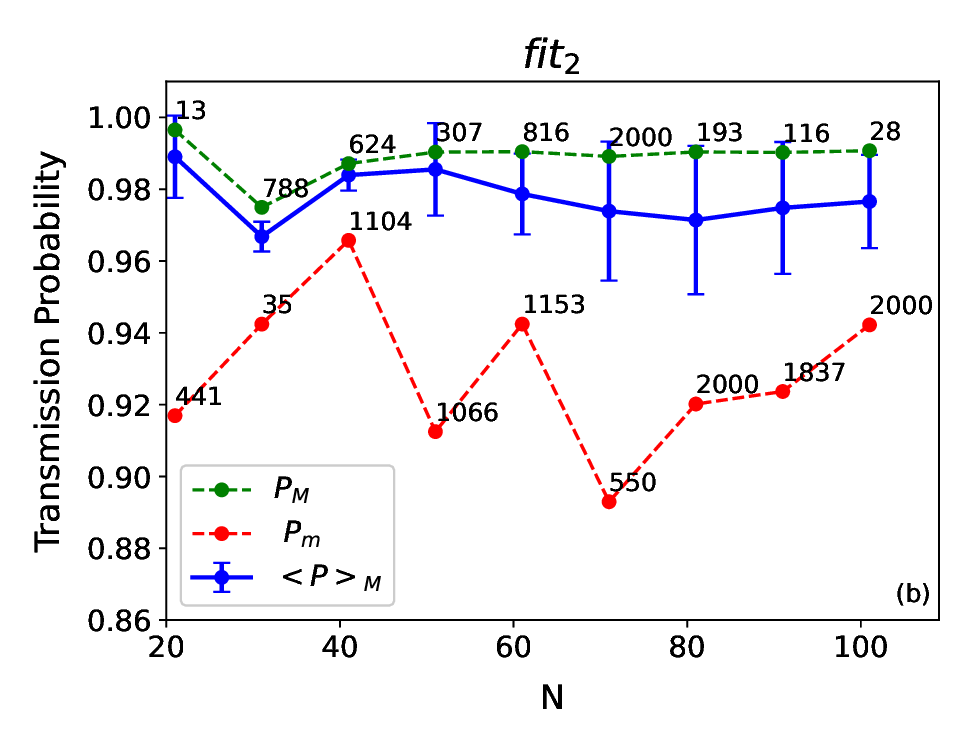}
\caption{The Figure shows the behavior of the maximal probability, Eq. \eqref{eq:maximal-probability}, the average transmission probability, Eq. \eqref{eq:average-trans-probability}, and the minimal probability, Eq. \eqref{eq:minimal-probability}, for chains with lengths ranging from 21 up to 101 spins, and for both fitness functions a) $fit_1$, b) $fit_2$. The green, blue, and red dots correspond to $P_M$, $\langle P \rangle_M$, and $P_m$, respectively. The data result of running the algorithm 50 times, from the random initialization of the population until one of the halting conditions is satisfied. The numerical value near the green and red points indicates in which generation the algorithm reaches that particular value of the corresponding probability. Table \ref{table:contorno1000} contains the hyperparameters used by the algorithm. The tolerance employed was 0.01.  The bar errors shown correspond to the standard deviation of the average transmission probability. In this Figure and the following ones, we used $\beta=0.9$ see Eq. \eqref{eq:fit2}. }\label{fig:transmission-probabilities}
\end{figure}

Figure~\ref{fig:transmission-probabilities} shows the transmission probability value found using the genetic algorithm, with the parameters shown in Table 1, for chains with different lengths, ranging from $N=21$ up to $N=101$. The algorithm runs for fifty independent experiments, from the initialization until it halts. The data included corresponds to the maximum value of the probability of transmission found for all the experiments and generations over which the algorithm runs. We look for the highest value with three restrictions, imposing a tolerance value such that if $P>0.99 $ the algorithm stops, or when the algorithm runs up to the maximum number of generations allowed, and if the maximum value for the transmission probability remains without change for 20 generations. The Figure also shows the minimum value recorded for the transmission probability and the average value. The "minimal" probability corresponds to the minimum value found in the populations of the last generations. Each run of the algorithm has a maximum value of the transmission probability, so we average these 50 values and plot them for each chain length. Also shown in the Figure are the generations at which the fitness attains each value. 

In other words, if $\nu$ is the index that numbers the runs of the algorithm, and $P_{1, N}^{\nu}(T)$ is the maximum value of the transmission probability of the fittest individual of any generation of the $\nu-$th run then, Figure \ref{fig:transmission-probabilities} shows the maximal probability 
\begin{equation}\label{eq:maximal-probability}
P_M = \max_{\nu} P_{1, N}^{\nu}(T).
\end{equation}
The average transmission probability is given by

\begin{equation}\label{eq:average-trans-probability}
\langle P \rangle_M = \frac{1}{N_r} \sum_{\nu=1}^{N_r}  P_{1, N}^{\nu}(T),
\end{equation}
where $N_r=50$, and the minimal probability is given by

\begin{equation}\label{eq:minimal-probability}
P_m = \min_{\nu}   p_{1, N}^{\nu}(T), 
\end{equation}
where $p_{1, N}^{\nu}(T)$ is the minimum value that attains the transmission probability at the last generation when the algorithm halts. We plot both quantities in Figure~\ref{fig:transmission-probabilities}, too. Table~\ref{table:contorno1000} shows the parameters used by the genetic algorithm to obtain the data in Figure~\ref{fig:transmission-probabilities}. 

\begin{table}[H]
\begin{center}
    
\begin{tabular}{|cc|c|cc|l}
\cline{1-2} \cline{4-5}
\multicolumn{2}{|c|}{Algorithm initialization}                  &  & \multicolumn{2}{c|}{Crossover and parent selection}  &  \\ \cline{1-2} \cline{4-5}
\multicolumn{1}{|c|}{Max number of generations} & 2000  &  & \multicolumn{1}{c|}{Number of parents}  & 200      &  \\ \cline{1-2} \cline{4-5}
\multicolumn{1}{|c|}{Population size}  & 1000 indiviuals   &  & \multicolumn{1}{c|}{Parent selection type} & sss      &  \\ \cline{1-2} \cline{4-5}
\multicolumn{1}{|c|}{Min value for a gene}       & 0                 &  & \multicolumn{1}{c|}{Elitism}          & 100      &  \\ \cline{1-2} \cline{4-5}
\multicolumn{1}{|c|}{Max value for a gene}       & $N$             &  & \multicolumn{1}{c|}{Crossover type}     & uniform &  \\ \cline{1-2} \cline{4-5}
\multicolumn{1}{|c|}{Saturation}                & 20 generations   &  & \multicolumn{1}{c|}{Probability}      & 0.6      &  \\ \cline{1-2} \cline{4-5}
\end{tabular}

\caption{The Table shows the hyperparameters used to obtain the data in Figure \ref{fig:transmission-probabilities} For the definitions of the size of the population, the maximum number of generations over which the algorithm runs, and the number of parents in the text, see the text. For the definitions of the other hyperparameters, see Appendix \ref{ap:genetic-algorithm}.} 
\label{table:contorno1000}
\end{center}

\end{table}

The three quantities, $P_M$, $\langle P \rangle_M $, and $P_m$, quantify different aspects of the searching algorithm, but all point to the same conclusion: the genetic algorithm consistently provides ECC that determine chains with very good to excellent transfer ability. 

Since the average transmission probability $\langle P \rangle_M$ values are very close to the maximum attainable values $P_M$ for all the chain lengths, as shown in Figure~\ref{fig:transmission-probabilities}, we conclude that the algorithm can easily reach transmission probabilities on the order of 0.98. Note that even the worst chains generated by the algorithm have transmission probabilities larger than 0.93, approximately. That these worst cases appear at the last generation indicates that running the algorithm for more generations would probably lead to better results.  

In those cases where the worst transmission probability appears in a generation that is not the last, it is clear that the algorithm becomes stuck and that the crossover and mutation rules can not modify the ECC of the individual chain. This condition happens because of the halting criteria that interrupt the corresponding run of the algorithm. 

\begin{figure}[bt]
\includegraphics[width=0.45\linewidth]{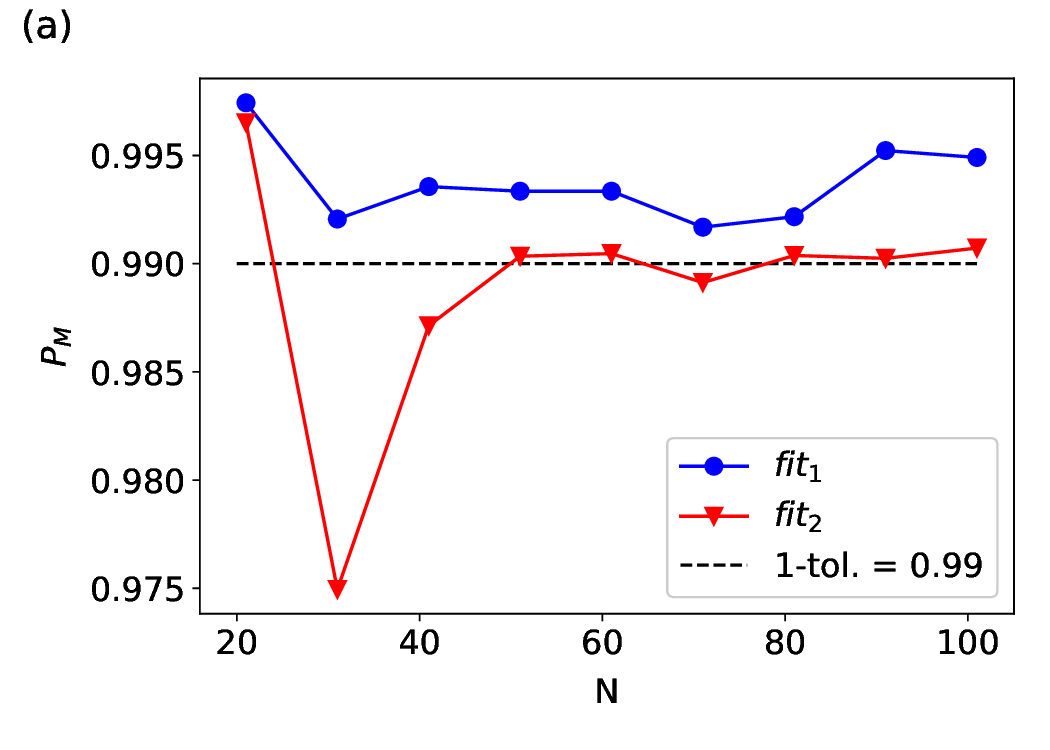}
\includegraphics[width=0.45\linewidth]{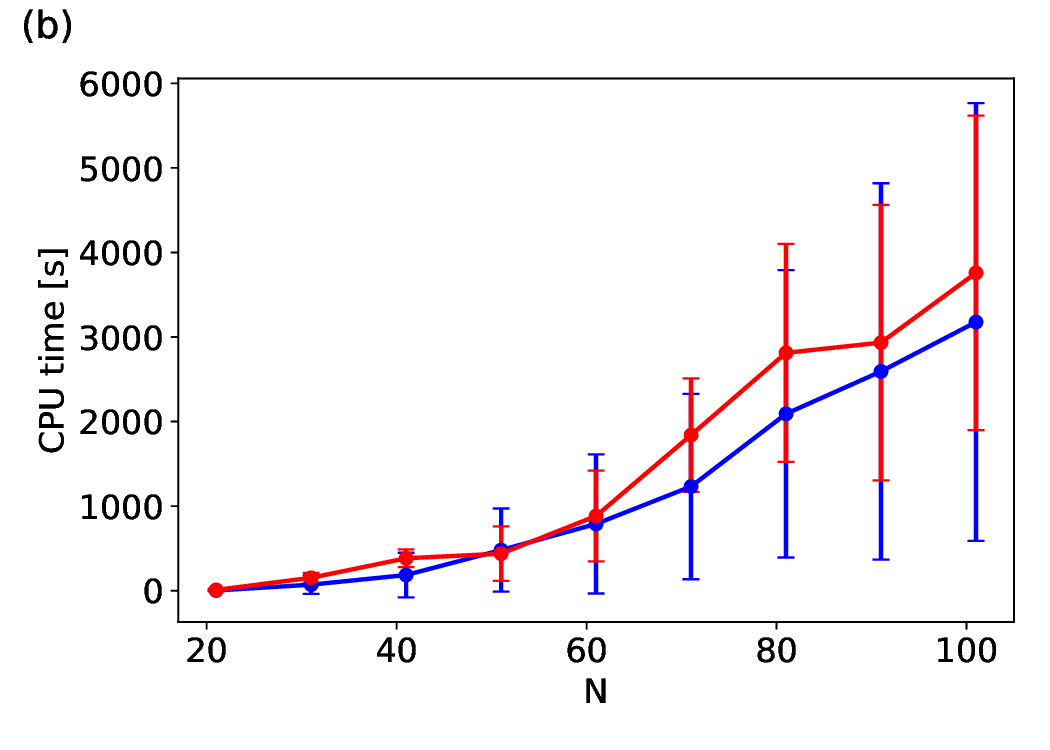}

\caption{The figure shows the maximal probability vs the chain length obtained for both fitness functions. The data points correspond to the green dots in both panels in Figure \ref{fig:transmission-probabilities}. The tolerance used by the genetic algorithm is 0.01. The algorithm always reaches the required tolerance when the fitness function is given by $fit_1$, while for the other fitness function, $fit_2$, the algorithm halts before attaining the tolerance. In this case, the algorithm needs a different set of hyperparameters than those in Table \ref{table:contorno1000} to achieve a maximal transmission probability bigger than 0.99.   }
\label{fig:performance-comparison}
\end{figure}

\begin{figure}[bt]
\includegraphics[width=0.3\linewidth]{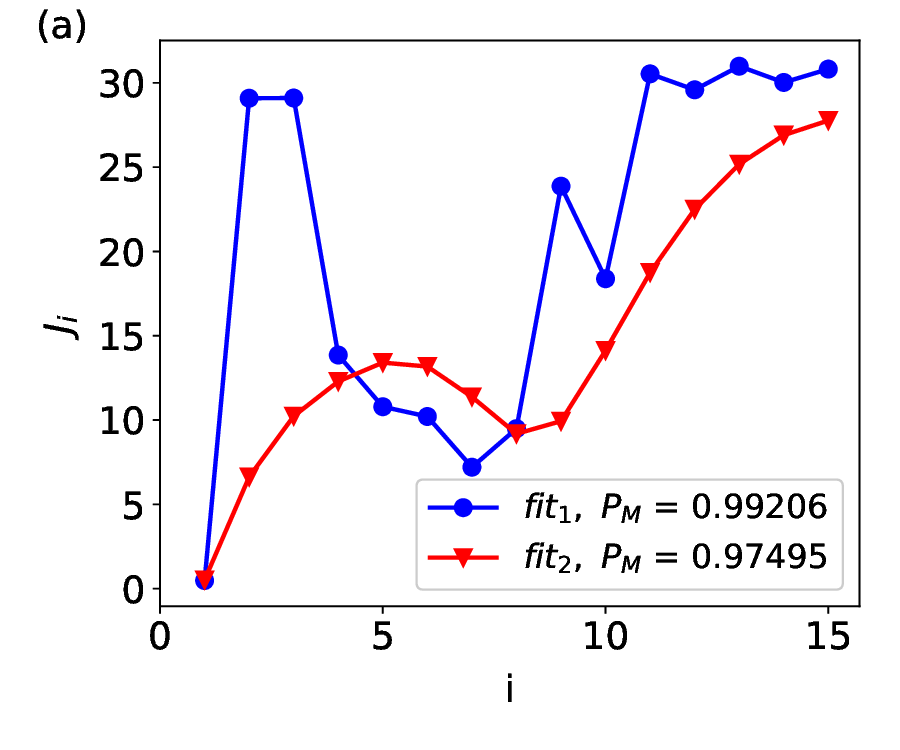} 
\includegraphics[width=0.3\linewidth]{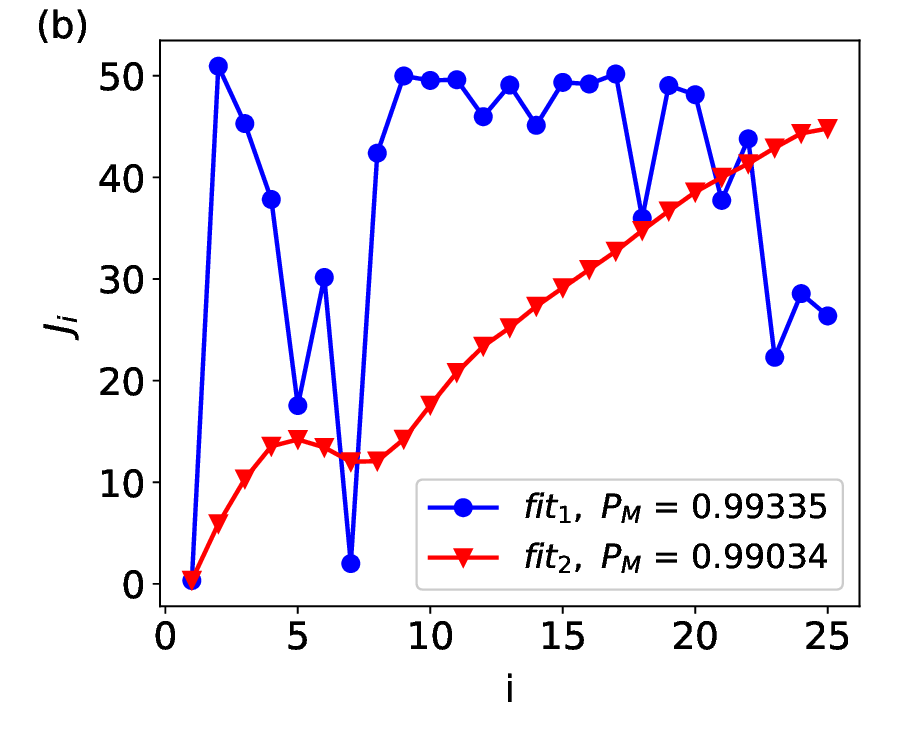} 
\includegraphics[width=0.3\linewidth]{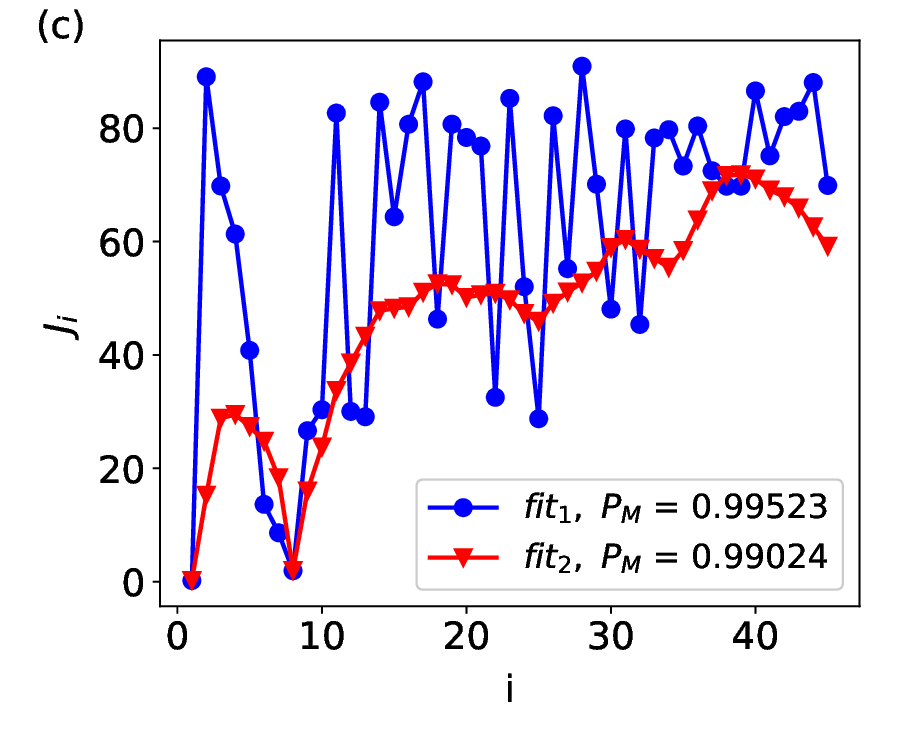} 
\caption{ The figure shows the strength of the exchange coupling coefficients of centrosymmetric chains obtained from the genetic algorithm vs the position index, $J_i \mbox{vs} \, i$, for chain lengths of a) $N=31$, b) $N=51$, and c) $N=91$. The set of the ECCs for a given chain is called an ECC distribution. The values shown correspond to chains obtained using the fitness functions, $fit_1$ and $fit_2$, and are shown with blue and red dots, respectively. The effect of the exponential modulation in $fit_2$ is evident. It reduces the distance between successive ECCs. The transmission probability values for each chain, at a fixed length, are comparable and shown in each panel. }
\label{fig:effect-of-b}
\end{figure}

Before changing the subject, it is worth comparing the values of $P_M$ obtained by running the algorithm with both fitness functions. In Figure~\ref{fig:performance-comparison} a), we plot both sets of data. The maximum transmission probability attainable with both fitness are quantitative and qualitatively similar. The computational costs are similar, although slightly bigger when the algorithm uses $fit_2$, see Figure~\ref{fig:performance-comparison} b). We also aim to compare the chains that result from optimizing the functions $fit_1$ and $fit_2$. As $fit_1$ does not impose any conditions over the ECC, other than the function must be optimized, it is not surprising that the ECC obtained for a given chain length seems random, or at least a rapidly changing function, see the set of blue points in Figure \ref{fig:effect-of-b}, which shows the ECCs for chains with three different lengths. The ECCs correspond to chains with good transfer ability with transmission probability $P>0.99$.

On the other hand, the green dots sets in Figure~\ref{fig:effect-of-b} correspond to ECCs obtained using $fit_2$. The difference between the performances of both types smoothed ($fit_2$) or rough ($fit_1$), is negligible when the tolerance imposed is equal to 0.01. 

As we will show, there are more similarities between smooth and rough chains. In the next Section, we will address the robustness of the transmission probability against static disorder. Nevertheless, we want to point out that the similarities strongly depend on the value of the tolerance imposed. We choose the factor in Eq.\ref{eq:fit2} that multiplies the transmission probability precisely to achieve the smoothhening of the ECC distribution without compromising the transmission probability. Nevertheless,  some loss is inevitable see Figure \ref{fig:performance-comparison}. 

\begin{figure}[bt]
\includegraphics[width=0.5\linewidth]{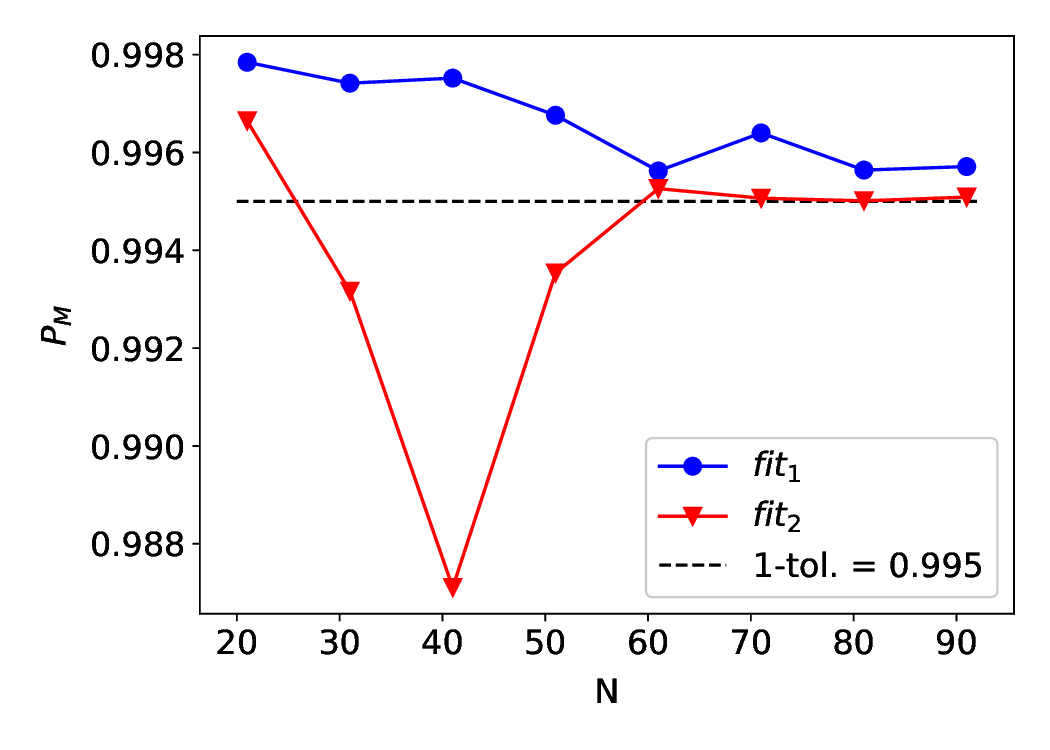}
\caption{The figure shows the maximal probability vs the chain length obtained for both fitness functions. The tolerance used by the genetic algorithm is 0.005. The algorithm always reaches the required tolerance when the fitness function is given by $fit_1$, while for the other fitness function, $fit_2$, the algorithm halts before attaining the tolerance. The set of hyperparameters used to obtain the data is the same  of Table~\ref{table:contorno1000}, except for the Max value for a gen that is set equal to $1.2\times N$, and the Max number of generations which is set to $3000$ }\label{fig:performance-comparison-2}
\end{figure}

Becoming greedy or because the task requires a stringent restriction could change the scenario completely. Figure~\ref{fig:performance-comparison-2} shows the results obtained, asking for a tolerance of 0.005.

Because the stringent tolerance could hinder the algorithm convergence, we employed a different set of hyperparameters. We used 3000 as the maximum number of generations, and the genes could vary between $\left[0, 1.2\times N\right]$, where $N$ is the chain size. The population size, number of parents, elitism, and other hyperparameters remained unaltered. 

\section{The effect of static disorder}\label{sec:static-disorder}

Each time a new method to obtain chains with good transfer abilities arises, it is necessary to analyze how much it is affected by the static disorder. The need for this analysis emerges because the nanoscopic or microscopic realization of the qubits of the chain would present differences from the ideally designed one. In other words, because of manufacturing defects, the interactions between the qubits do not have the design values but others. 

One of the disorder models more often analyzed considers that each interaction changes proportionately to its design value 
\begin{equation}
J_i \longrightarrow J_i  (1+ \xi_i),
\end{equation}
where $\xi_i$ is a random gaussian variable  with null mean and standard deviation $\sigma$. The random variables affecting each ECC could be, or not, correlated. Note that this disorder model is quite similar to a mutation, except that in the version of the genetic algorithm employed, a mutation keeps the centrosymmetric character of the chain. In contradistinction,  the disorder effectively breaks this symmetry. Another difference lies in the number of ECCs affected by a mutation and those affected by a realization of the disorder. The disorder affects all the ECCs. The mutation does not. 

There are several quantities worth studying to characterize the effect of static disorder. Starting with an ECC distribution with a high transmission probability, we calculate the mean transmission probability, which is given by

\begin{equation}\label{eq:mean-transmission-probability}
\langle P_{1,N}(T) \rangle_{\xi} = \frac{1}{N_{\xi}} \sum_{\kappa=1}^{N_{\xi}} P_{1,N}(\lbrace\xi\rbrace_{\kappa},T), 
\end{equation}

where $N_{\xi}$ is the number of realizations of the static disorder, and $P_{1, N}(\lbrace\xi\rbrace_{\kappa}, T)$ is the transmission probability obtained with the ECC distribution corresponding to the $\lbrace \xi \rbrace_{\kappa}$ realization of the random variables $\xi_i$. A realization of the random variables $\xi_i$ is a set of outcomes sorted accordingly with the probability distribution of the random variables. 

\begin{figure}[bt]
\includegraphics[width=0.3\linewidth]{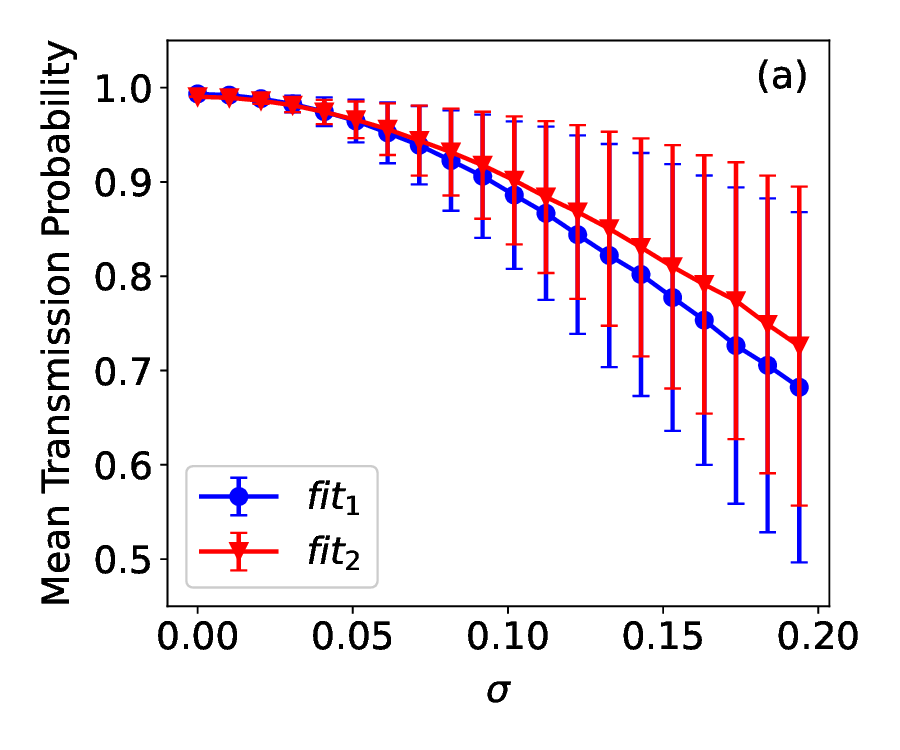}
\includegraphics[width=0.3\linewidth]{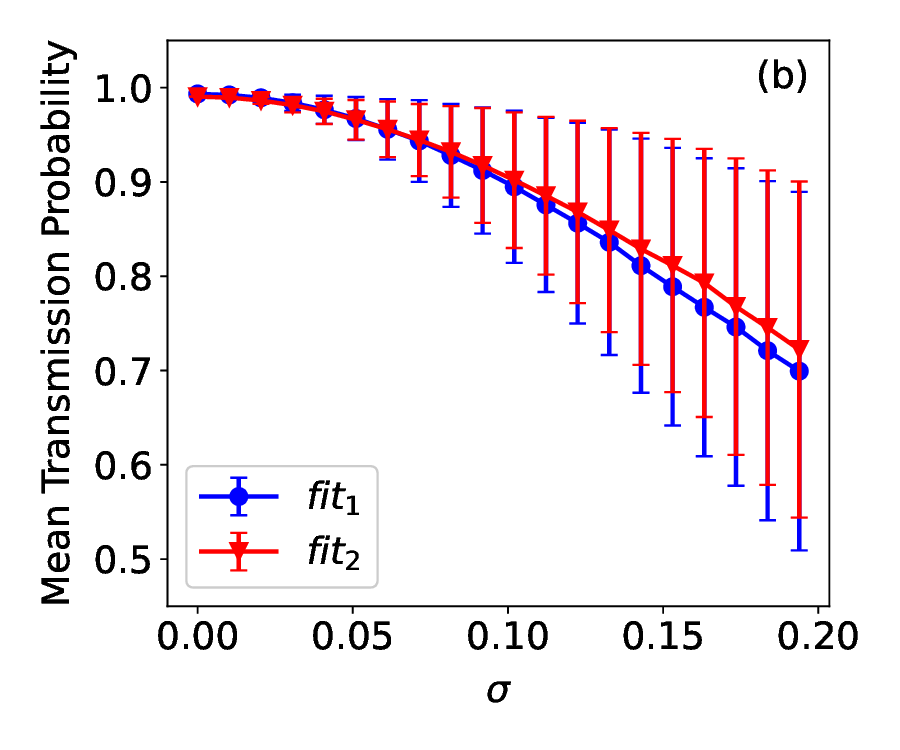}
\includegraphics[width=0.3\linewidth]{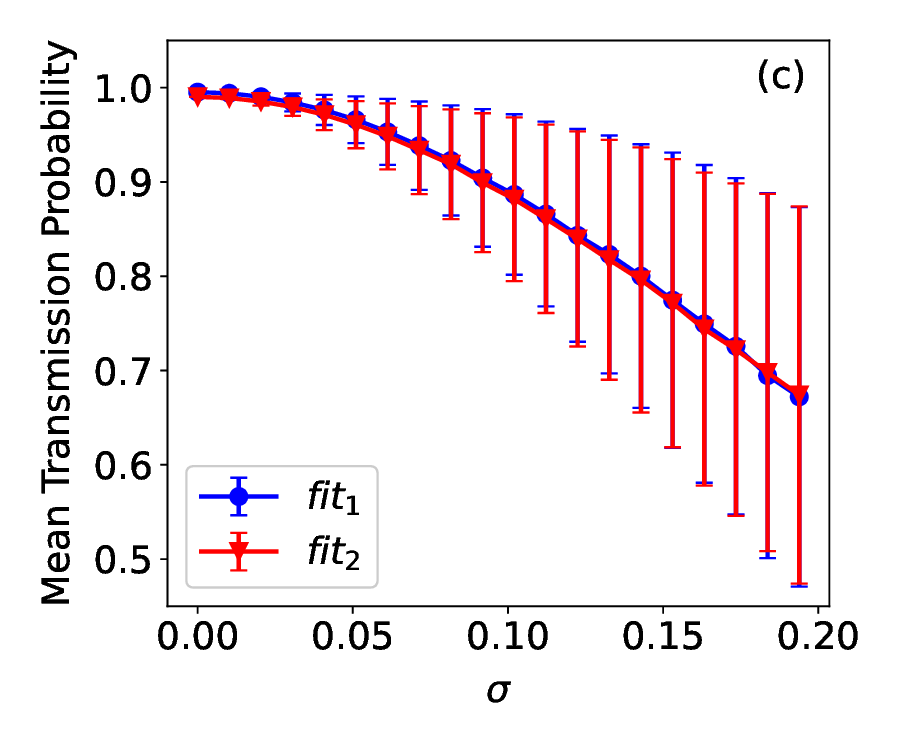}
\caption{The figure shows the mean transmission probability and its standard deviation, Eqs. \eqref{eq:mean-transmission-probability} and \eqref{eq:variance-mtp}, vs the disorder strength $\sigma$, obtained for both fitness functions and three chain lengths, a) $N=51$, b) $N=61$, and c) $N=91$. Note that the decaying behavior of the mean transmission probability is qualitatively the same, irrespective of the EEC distribution type. Also, note that the transmission probability for $\sigma=0$ is comparable for all the cases analyzed. The error bar size corresponds to the value of the standard deviation, Eq.~\eqref{eq:variance-mtp}. }\label{fig:static-disorder}
\end{figure}

\begin{figure}[bt]
\includegraphics[width=0.45\linewidth]{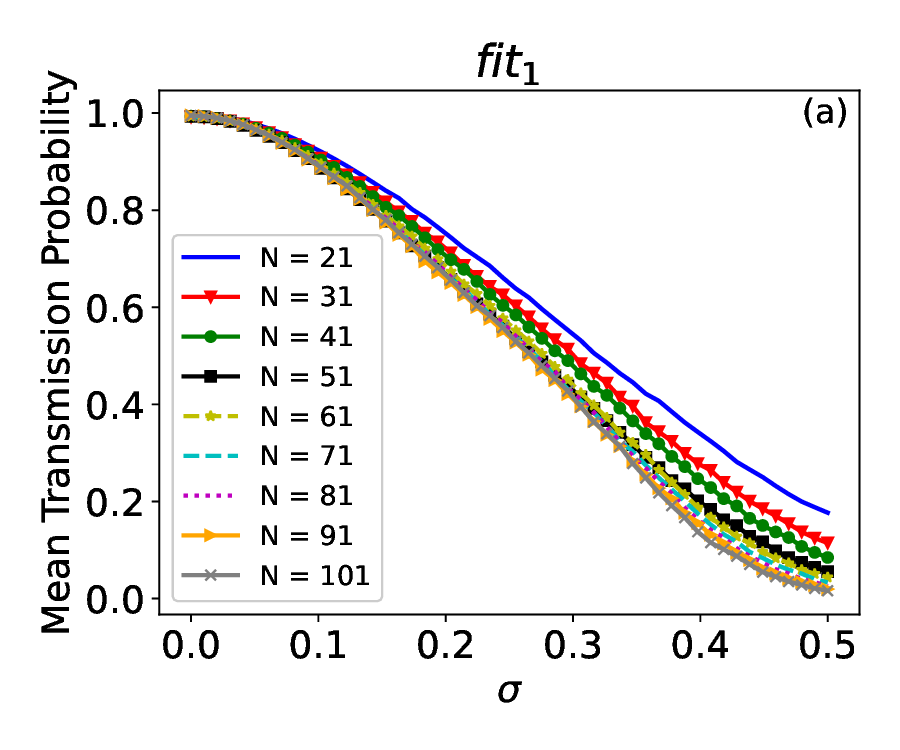}
\includegraphics[width=0.45\linewidth]{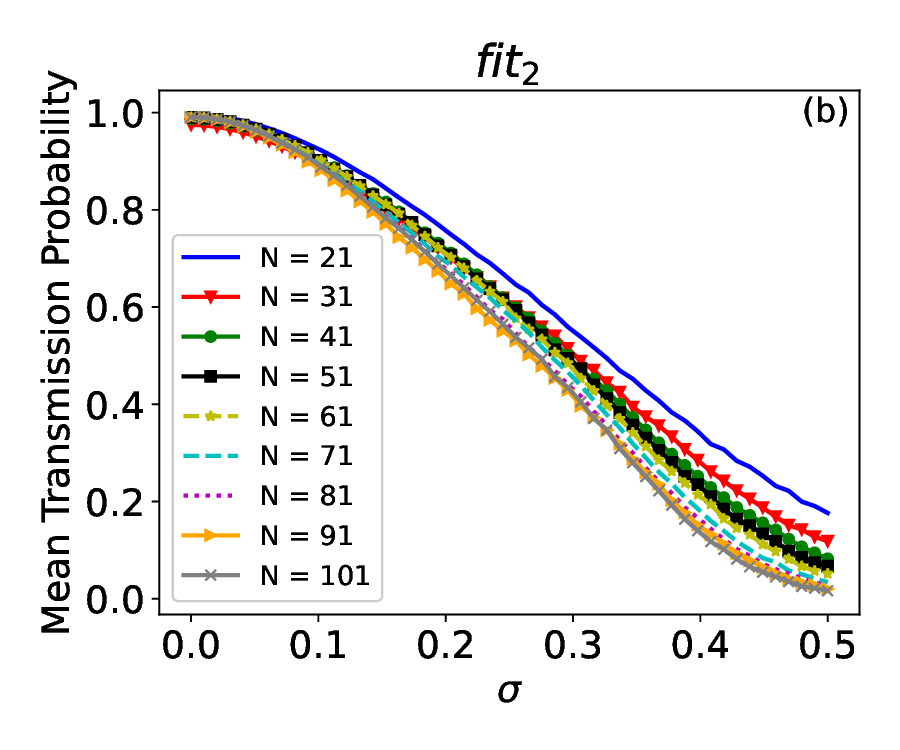}
\caption{The figure shows the mean transmission probability vs the disorder strength $\sigma$, obtained for both fitness functions, a) $fit_1$, b) $fit_2$, and nine chain lengths. }\label{fig:meanPvsN}
\end{figure}

Since we have two kinds of EEC distributions, softened and rough, we want to compare their robustness against static disorder. Figure \ref{fig:static-disorder} shows the mean transmission probability obtained for chains with both kinds of ECC distributions. The unperturbed EECs correspond to chains with transmission probabilities larger than 0.99. The three panels show the mean transmission probability for chain lengths of $51,61$ and $91$ spins against the disorder strength $\sigma$. Interestingly, the decay of the mean transmission probability is independent of the ECC type and the length $N$. The bar errors correspond to the standard deviation of the transmission probability,
\begin{equation}\label{eq:variance-mtp}
\sigma_P = \sqrt{\frac{1}{N_{\xi}} \, \sum_{k=1}^{N_{\xi}} \, \left(
\langle P_{1,N}(T) \rangle_{\xi} - P_{1,N}(\lbrace\xi\rbrace_{k},T)
\right)^2 }.
\end{equation}
The data in the Figure~\ref{fig:meanPvsN} shows that the mean transmission probability is almost independent of the chain length and type. Chains with ECC distribution rough and softened have the same mean transmission probability for a given chain length and disorder strength.  

On the other hand, the chains with softened ECC distributions have lesser variance values for fixed length chain and disorder strength, although this advantage is less and less noticeable for larger values of $N$. 

That the mean transmission probability depends sligthly on the chain length is attributable to the small number of eigenstates and eigenvalues that contribute to the QST. Only the eigenvectors whose eigenvalues satisfy the so-called Kay's conditions \cite{Kay2010}  contribute effectively to the state transfer, and their number grows very slowly with the chain length, see reference \cite{FerronPS}. 

 It is worth analyzing carefully the dependency of the mean transmission probability with the chain length. Figure \ref{fig:meanPvsN} shows the behavior of the mean transmission probability for different chain lengths and both fitness functions.

In this Figure, we plot the mean transmission probabilities for values of the disorder strength $\sigma$ up to $0.5$. Such large values of static disorder are not realistic but allow us to study the decay of the transmission probability from near the unity to values close to zero.  

Both panels in Figure \ref{fig:meanPvsN} show that the mean transmission probability effectively depends on the chain length and the fitness function chosen, but the dependency is smooth.

 We fitted an exponential function to each curve shown in both panels to analyze the dependency with the chain length. The fitting function is given by
\begin{equation}\label{eq:fit-mean}
P_f = a(N) \exp{(-b(N) \sigma^{c(N)}) }+ d(N) ,
\end{equation}
where $a(N),b(N),c(N)$ and $d(n)$ are the parameters of the fit. The Gaussian or almost Gaussian behavior of the mean transmission probability has been identified previously in XX chains with perfect transmission and perturbed by random magnetic site-dependent fields by De Chiara et al. \cite{DeChiara2005}, and in XX chains with optimized interactions and static disorder by Zwick et al. \cite{Zwick2012}.

\begin{figure}[bt]
\includegraphics[width=0.45\linewidth]{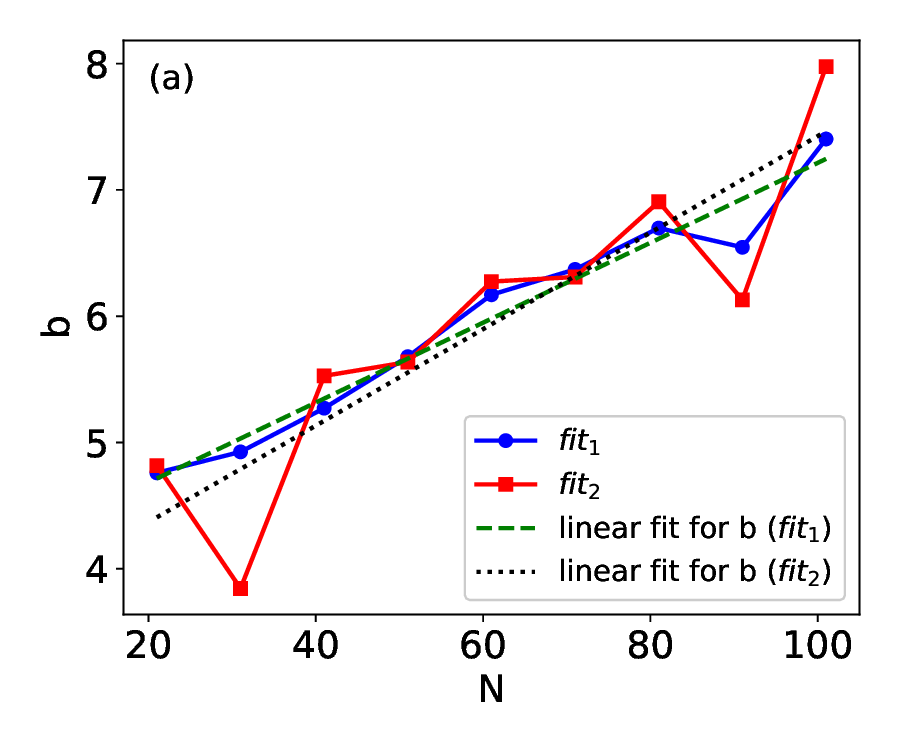}
\includegraphics[width=0.45\linewidth]{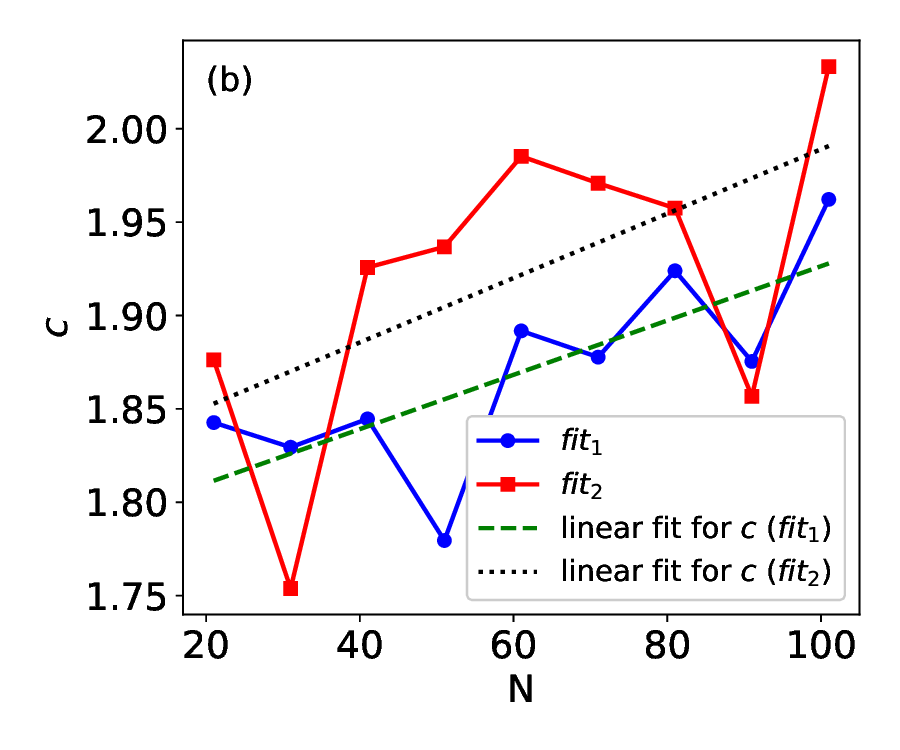}
\caption{ The Figure shows the values for two of the fitting coefficients defined in Equation~\ref {eq:fit-mean} for different values of the chain length,  a) shows $b(N)$, and b) shows $c(N)$.  Data points correspond to chains obtained with both fitness functions, the blue ones to $fit_1$ and the red ones to $fit_2$, respectively.  The dashed lines in panel a) are linear fits of both sets.  Except for the chains with lengths $N=31$ and $N=91$ obtained with the fitness function $fit_2$, the agreement between the fits and data points in panel a) is excellent, the disagreement for the chains with $N=31$ and $N=91$ attributable to their low transmission probability for zero disorder when compared with the other chains. The data point values in panel b) show that for most cases, the exponent $c(N)\leq 2$.}\label{fig:fitting-coefficient}
\end{figure}

The coefficient $b(N)$ is proportional to the chain length $N$ in both References. In Figure~\ref{fig:fitting-coefficient}, we plot the coefficient $b(N)$ obtained by fitting the curves shown in Figure~\ref{fig:meanPvsN}. 

Figure~\ref{fig:fitting-coefficient} a) shows the behavior of $b(N)$ versus the chain length $N$. As we fitted the mean transmission probability for both types of chains, there are two sets of data points labeled $fit_1$ and $fit_2$. The dashed lines correspond to linear fits of each set of points. The fits support that $b(N)$ has a linear dependency with the chain length, especially for the chains obtained with the fitness function $fit_1$. The smallness of the slope of both dashed lines is consistent with the behavior observed in Figure~\ref{fig:meanPvsN}. The mean transmission probabilities for larger chains do not decay much faster than the probabilities of the shorter chains.

Figure~\ref{fig:fitting-coefficient} b) shows the behavior of $c(N)$ versus the chain length $N$. The data values for both types of chains are consistent with an almost Gaussian dependency of the mean transmission probability with the disorder strength.  

The dashed lines in Figure~\ref{fig:fitting-coefficient} b)  correspond to linear fits of both sets and are included as a guide to the eye.

\section{The perspective from the random matrix theory}\label{sec:random-matrices}

That the effect of static disorder over both chain types investigated so far is pretty much the same is, to some extent, surprising. A stronger dependency on the decaying rate with the chain length would signal that the mean transmission probability depends on global properties of the Hamiltonian, as the whole spectrum and their eigenvalues. While Eq. \eqref{eq:transmission-probability} indicates that that is the case, it is clear that not all the eigenvectors have the same weight. 

Following Kay \cite{Kay2019}, near-perfect QST can be expected in Heisenberg
spin
chains whose spectrum satisfies $E_{i+1} - E_{i} \sim q_i \alpha$, where $q_i$
is
an odd natural number, {\em i.e.} successive energy values should differ in odd
multiples of a constant $\alpha$, approximately. Optimization methods produce spectra with only a handful of successive eigenvalues that satisfy Kay's condition \cite{SerraPLA, FerronPS}. We termed this scenario a partially ordered spectrum.  

The following paragraphs present an argument discussing the consequences of having a partially ordered spectrum. 

The time-dependent quantum state of the chain is given by
\begin{equation}\label{eq:psi-decomposed}
|\psi(t)\rangle = \sum_{\tilde{i}} \chi_{\tilde{i}} \exp{\left( -i E_{\tilde{i}}
t\right)} |v_{\tilde{i}}\rangle
+ \sum_{i\neq\tilde{i}} \chi_{i} \exp{\left(-i E_{i} t\right)} |v_i\rangle,
\end{equation}

\noindent where the index $\tilde{i}$ runs over the eigenvalues that satisfies
$\Delta_{\tilde{i}}/(\pi/T)\sim q_i$, then at $t=T$ the terms on the first sum
will
interfere constructively. The vectors $|v_i\rangle$ are
solutions of
\begin{equation}
 h_N ({J}_T) |v_i\rangle = E_i |v_i\rangle .
\end{equation}

The initial state 
\begin{equation}
|\psi(t=0)\rangle = |\mathbf{1}\rangle = \sum_{\tilde{i}} \chi_{\tilde{i}}  |v_{\tilde{i}}\rangle
+ \sum_{i\neq\tilde{i}} \chi_{i}  |v_i\rangle,
\end{equation}
allow us to calculate the coefficients
\begin{equation}
 \chi_i = \langle v_i | \mathbf{1}\rangle.
\end{equation}

It is convenient introducing the coefficients 
$c_i= |\chi_i|^2= |\langle v_i | \mathbf{1}\rangle|^2$.

Near perfect QST occurs at arrival time $T$ when a
number of terms in Eq.~\eqref{eq:psi-decomposed} interfere constructively
to
produce $|\langle \psi(T)| N\rangle |^2 \sim 1$.
The relatively reduced number of eigenvalues that
satisfies approximately the constructive interference condition

\begin{equation} \label{eq:constructive-interference}
 \Delta_i/(\pi/T)= q_i
\end{equation}

\noindent indicates that it is necessary to get
$\chi_{i}\sim 0$ for $i\neq\tilde{i}$  to obtain near-perfect QST. Note that the number of eigenvalues that approximately satisfies the constructive interference condition does not change much when the chain length increases.

Consequently, the initial state is a superposition of those eigenvectors whose energies satisfy $\Delta_i = q_i (\pi/T) $, while the other eigenvectors contribute negligibly to the superposition. The eigenvectors corresponding to the
almost ordered eigenvalues show strong localization at both extremes of the
quantum chain. Moreover, since 
\begin{equation}\label{eq:mirror-coefficients}
|\langle\mathbf{1}|v_i\rangle| = |\langle\mathbf{N}|v_i\rangle|,
\end{equation}
the condition that $|\psi(t=0)|^2 \sim 1$ ensures that $|\psi(t=T)|^2 \sim 1.$  

In a few words, almost near perfect QST is achievable with a reduced number of
eigenvalues satisfying condition \eqref{eq:constructive-interference} as long
as their eigenvectors are well-localized
at the extremes of the spin chain. What our results indicate is that the genetic algorithm finds precisely the ECC distribution that results in the almost ordered spectrum with well-localized
eigenvectors that are required for the almost perfect QST to happen.

The conditions in Eq, \eqref{eq:constructive-interference} and \eqref{eq:mirror-coefficients} ensure that the transmission probability of a chain with optimized ECC should be a quasi-periodic function of time but, eventually, the terms in the second sum in Eq. \eqref{eq:psi-decomposed} blur the periodic-like behavior making the analysis of the time-dependent transmission probability difficult. So, instead of analyzing the dependency on time, we will study the statistical properties of the spectra of the Hamiltonians of chains designed using our genetic algorithm. 

There are several quantities used to analyze the spectrum of many-body Hamiltonians, mainly the distribution of the spacings between neighboring energy levels, see reference \cite{Gubin2012} and references therein, and the consecutive-gap ratio $r$ \cite{Oganesyan2007, Atas2013, Siegl2022} defined as
\begin{equation}
r_n = \frac{\min{\lbrace s_n, s_{n-1}\rbrace}}{\max{\lbrace s_n,{s_n-1}\rbrace}}
\end{equation}
where $s_n = E_{n+1} - E_n$ is the difference between neighboring energy levels, each corresponding to a multiplet of a given total spin quantum number.

\begin{figure}[bt]
\includegraphics[width=0.4\linewidth]{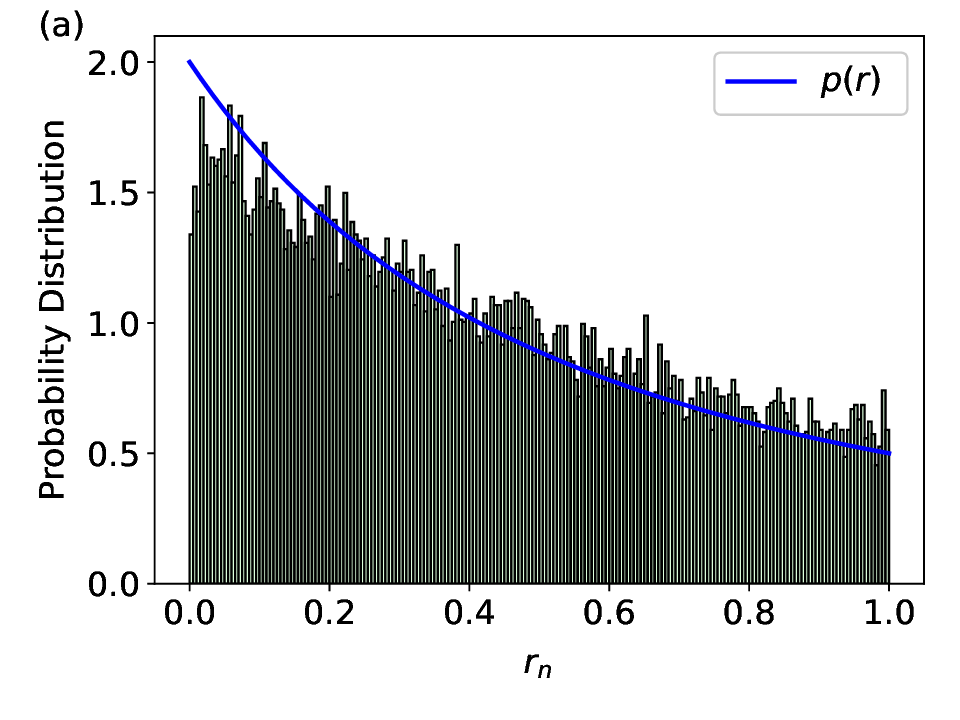}
\includegraphics[width=0.4\linewidth]{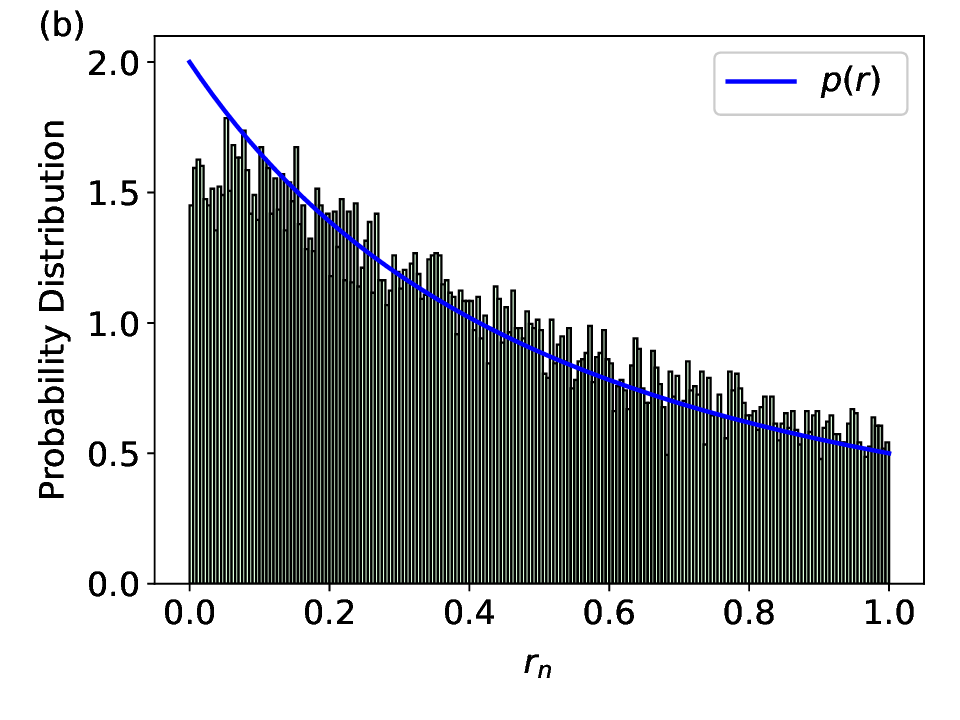}
\includegraphics[width=0.4\linewidth]{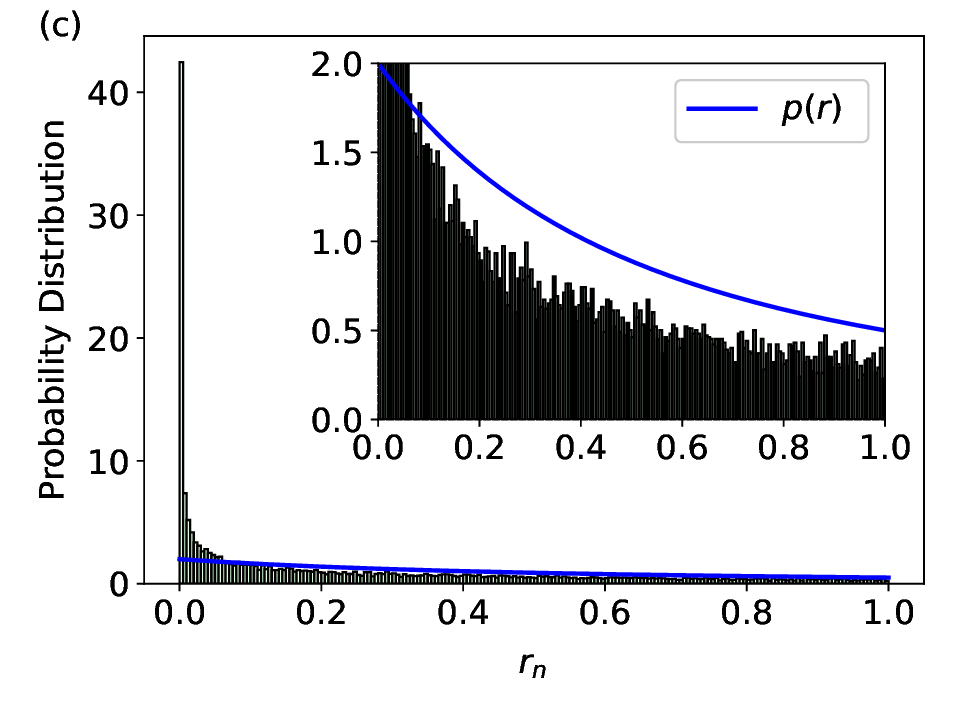}
\includegraphics[width=0.4\linewidth]{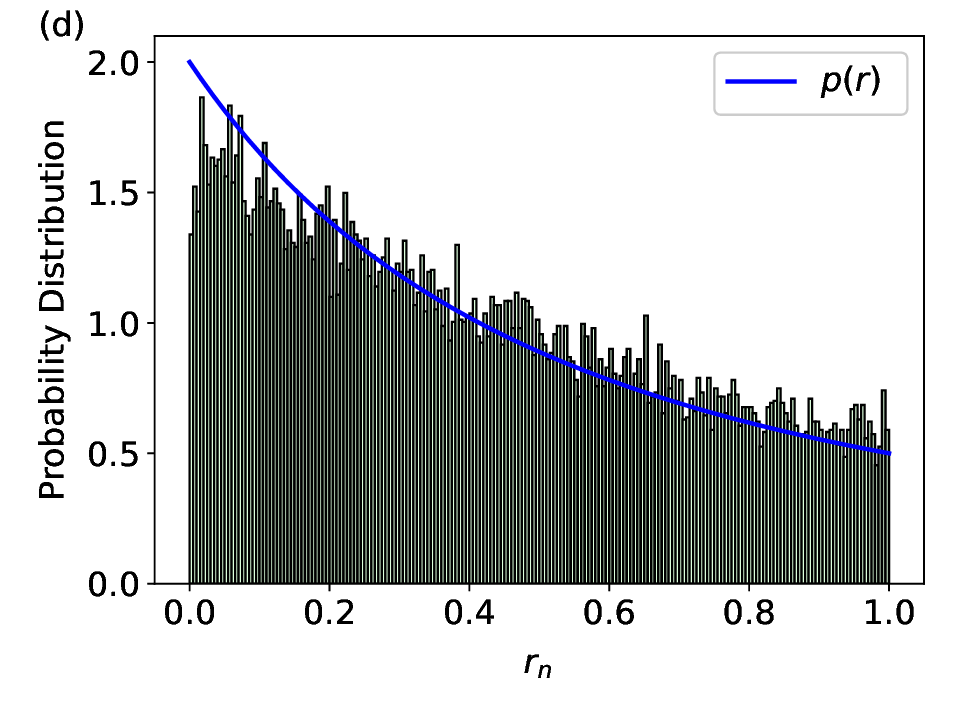}
\caption{The figure shows the probability distribution for the consecutive gap ratio $r_n$ obtained for site-dependent designed chains.  The histograms in the four panels show the consecutive gap ratio values obtained for chain Hamiltonians obtained using our algorithm with the two fitness functions and for two different lengths.  The interval $\left[0,1\right]$ accommodates two hundred bins, and the height of the histogram in each bin corresponds to the number of consecutive gap ratios in the bin interval normalized by the total number of $r$ values. The chain length and fitness function corresponding for each panel are: a) $N=21$, $fit_1$, b)$N=21$, $fit_2$, c) $N=51$, $fit_1$, and d) $N=51$, $fit_2$. Note that the eigenvalues used to calculate the consecutive gap ratios include all the eigenvalues from the subspaces from one up to five excitations when $N=21$ and up to three when $N=51$. The blue line depicts the probability distribution $p(r)$ of a Hamiltonian with Poissonian spacing levels.}\label{fig:gap_ratio}
\end{figure}

The mean value and distribution of the consecutive gap ratio for random matrices belonging to the different ensembles are well known. Both quantities distinguish Hamiltonians with Poissonian statistics from, for instance, chaotic statistics. In Figure~\ref{fig:gap_ratio}, we plot as histograms the values obtained for the consecutive gap ratio for four different chains, two obtained using $fit_1$ and two obtained with $fit_2$. The necessary eigenvalues were calculated using a standard package \cite{quspin}. The blue continuos curve shown in the four panels corresponds to the
probability distribution 
\begin{equation}
p(r) = \frac{2}{(1+r)^2} ,
\end{equation}
that is the distribution function for the consecutive gap ratio of a Hamiltonian showing an integrable phase \cite{Oganesyan2007, Siegl2022} whose level spacing is Poissonian. 

The four examples in Figure~\ref{fig:gap_ratio} show two trends shared by all the chains analyzed in this work. Panels b) and d) show that the consecutive gap ratios obtained from the eigenvalues of the Hamiltonians of chains constructed using $fit_2$ are consistent with the probability distribution of a Hamiltonian with Poissonian level spacing. Short chains obtained using $fit_1$ also show this property see Figure~\ref{fig:gap_ratio} a). For longer chain lengths, the histograms of consecutive gap ratios depart from the probability distribution associated with a Hamiltonian with Poissonian level spacing and develop a peak near zero, see Figure~\ref{fig:gap_ratio} b).

\section{Discussion and Conclusions}\label{sec:conclusions}

Since homogeneous Heisenberg chains do not present perfect transmission without a site-dependent external magnetic field, and the arrival times associated with {\em pretty good transmission} are excessively long, it seems clear that the most efficient way to achieve high-quality quantum state transfer without resorting to time-dependent external control is the use of site-by-site engineered chains. 

As the results presented in this work show, the genetic algorithm provides different solutions depending on the constraints imposed on the particular fitness functions employed.  Despite the roughness or softness of the exchange coupling coefficient distributions obtained, the physical properties of both types seem the same.  So, at least from the theoretical point of view, there is not a clear-cut better choice, although for implementing a transfer chain, the softer versions could be more appealing. 

It is worth discussing the effect of the exponential term depending on the differences between consecutive exchange coupling coefficients included in the expression of $fit_2$, Eq. \ref{eq:fit2}, since we only included results obtained fixing $\beta=0.9$.  The main effect is the softening of the ECC distribution. 

Choosing values of $\beta$ between 0 and 0.9 produces increasingly soft ECC distributions until the results become indistinguishable for $\beta \geq 0.9$.  The resulting chains show qualitatively and quantitatively the same transmission properties, so we show a single representative case. 

The scaling with the chain length in Eq. \ref{eq:fit2} also was analyzed.  Because the maximum attainable value for each exchange coupling coefficient is $O(N)$, it seems reasonable to choose that the difference between two successive couplings should be on this order, too.

The site-by-site chains designed are robust, and this property holds for all the lengths studied since any chain will transmit at a $98\%$ efficiency of its peak design value for a disorder strength of $5\%$, on average.  These figures correspond to the transmission probability, which results in a better performance for the fidelity of transmission of arbitrary quantum states.

For threshold values of $0.99$ ro $0.995$, the execution time of the algorithm roughly scales as $N^2$, where $N$ is the chain length, leading us to conclude that the genetic algorithm is quite efficient in finding ECC distributions that allow for high-quality quantum state transfer, at least when the transmission probability asked for is not too close to unity. 

Given that the fastest algorithms to calculate the eigenvalues and eigenvectors of a tridiagonal matrix are $O(N \log N)$, it is clear that the running time of the genetic algorithm should be  $O(N)$, at least for the examples analyzed.  Two-dimensional spin arrays with first-neighbor interactions would be optimizable without such a high computational cost if this scaling is confirmed,

The behavior of both coefficients shown in Figure~\ref{fig:fitting-coefficient} supports that chains with rough EEC distributions are more robust under static disorder when the longer the chain is, although the differences are not necessarily too significant. Both fitness functions produce ECC distributions whose Hamiltonians show localized eigenstates with their eigenvalues satisfying  Kay's conditions.  Localized eigenstates at the extremes of the chain are one of the more recognizable traits that the chain Hamiltonian must show to achieve high-quality quantum state transfer, and they are present in proposals involving topological states in SSH chains \cite{Estarellas2017}, tunable XXZ chains \cite{Serra2022}, and chains with long-range interactions. 

Even though all the examples mentioned above have localized eigenstates, it is worth noting that transmission in SSH-type chains takes place in the asymptotic regime, so the transmission times are enormous. 

It would be interesting if, appropriately choosing a fitness function, a transmission chain could be obtained such that it transmits at short times, with high quality, with localized states, and with some set of isolated states in the gap between bands.  

The genetic algorithm implemented in this work works effectively on different kinds of Hamiltonians with minimal adaptations.  We focused on the isotropic Heisenberg Hamiltonian because of its relevance to the area and the difficulty of analyzing it beyond the homogeneous chain and other variants with analytical solutions. Nevertheless, we think that the best application for our algorithm is in the context of systems with tunable site-dependent interactions, in particular chains composed of transmons and other types of superconducting qubits.

\vspace{30mm}

\noindent {\bf Acknowledgements}

OO and SPS acknowledge partial financial support from CONICET (PIP 11220210100787CO) and SECYT-UNC. All the authors acknowledge the support of the High-Performance Computing Center of the National University of Córdoba, Argentina (CCAD-UNC). 

\appendix

\section{Genetic Algorithm}\label{ap:genetic-algorithm}

In order to apply genetic algorithms to optimize transmissions in these systems, we define the chromosomes as the complete set of ECCs that represent a particular chain. Correspondingly, every single ECC constitutes a gene. Since the studied chains are center-symmetric, optimization is done over only half the ECCs.

Pygad library uses classes to run genetic algorithms. Each run is defined as an instance of a class and offers the possibility of tuning a wide variety of hyperparameters (class attributes). The ones used in our implementation were: 

\begin{itemize}

    \item \texttt{num\_generations} = Maximum number of generations over which the algorithm iterates, {\em i.e.} the maximum number of new populations generated.
    
    \item \texttt{fitness\_func} = Fitness function used to characterize the best solutions. Throughout this work, we discuss the implementation of two different fitness functions based in different attributes of the solutions, both of them are naturally based in the transmission probability.

    \item \texttt{sol\_per\_pop} = Total number of chromosomes in the population,  {\em i.e.} number of possible sets of ECCs for which fitness is calculated in every generation.
      
    \item \texttt{num\_parents\_mating} = Number of parent chromosomes selected to give origin to the next generation.
    
    \item \texttt{parent\_selection\_type} = Parent selection method. The chosen method is \textit{steady state selection} ('sss'). In this method, the best solutions (those with higher fitness) are chosen as parents and the worst solutions are replaced with the generated offspring. This ensures that a part of the population survives and pass on to the next generation.
    
    \item \texttt{num\_genes} = Number of genes in each chromosome. In this case, the number of genes is equal to half the number of ECCs. 
    
    \item \texttt{init\_range\_low/high} = Initialization range. Genes for each chromosome in the population are generated randomly within these limits. It is possible to substitute these parameters with the attribute \texttt{initial\_population}, in order to use a specific initial population.

    \item \texttt{keep\_elitism} = This parameter sets a fixed number of high-fitness valued solutions to keep for the next generation. 
    
    
    \item \texttt{crossover\_type} = Crossover method. We use the uniform crossover technique, which works by randomly selecting one of the 2 mating parents, copying the gene from it and assigning it to the same position in the resulting chromosome.
    
    \item \texttt{mutation\_type} =  Mutation type. We designed a custom mutation function (Algorithm \ref{alg:mutation}) that applies changes that are proportional to the value of each gene. The value of the selected gene ($J_i$) mutates according to $J_i \to J_i \ (1 + \delta) $, where $\delta$ is a random value chosen from a uniform distribution in the interval $[-\zeta,\zeta]$. We consider three different values of $\zeta$, which define the maximum possible strength of the mutations. Before the algorithm converges (new values are not similar to the mean of the last 10 values), the mutations should be stronger and we use a “strong mutation value”. Also, since the first gene varies in a wider range, we use a differentiated $\zeta$ value for this gene. After convergence, we use a “weak mutation value“. In this stage, the same value is used for every gene.

    \item \texttt{mutation\_probability} = Probability of a gene being affected by mutation
       
\end{itemize}

Once each of these parameters is configured, a method of the associated class called  \texttt{run}  allows the execution of the generated instance. The corresponding results are then saved as attributes of the same instance. Note that only the chosen methods for crossover, parent selection and mutation were described, the rest of the preset PyGAD functions for each of them can be found in the library documentation \cite{PyGAD}.

\begin{algorithm}[H]
\caption{Adaptive Mutation}\label{alg:mutation}

\begin{algorithmic}

\IF {$n_{generations}  < 10$} \hspace{5 pt}  \COMMENT{No convergence $\implies$ Strong Mutation} 

\STATE $J_i = J_i * (1 + \delta)  $  \COMMENT{$\delta \in [-\zeta,\zeta] \ \, \zeta = 0.05 \ (0.1 \text{if i = 0} ) $  }
\ELSE 
\STATE history = vector containing the last 10 fitness values
\STATE mean = $\sum \text{history} / 10$
\STATE $\Delta $ = history - mean
\IF { $ \ \exists \ d_j \in \Delta \textbf{  such that  }  d_j > 0.001 $}  \COMMENT{No convergence $\implies$ Strong Mutation}

\STATE $J_i = J_i * (1 + \delta)  $  \COMMENT{$\delta \in [-\zeta,\zeta] \ \, \zeta = 0.05 \ (0.1 \text{if i = 0} ) $  }

\ELSE \hspace{5 pt} \COMMENT{Convergence $\implies$ Weak Mutation} 
    \STATE $J_i = J_i * (1 + \delta)  $  \COMMENT{$\delta \in [-\zeta,\zeta] \ \, \zeta = 0.03 $  }
\ENDIF
\ENDIF

\end{algorithmic}
\end{algorithm}

\appendix

\end{document}